\RequirePackage{fix-cm}
\documentclass[twocolumn]{svjour3}          
\smartqed  
\usepackage{graphicx}
\usepackage{mathtools}
\begin{document}

\title{Efficient and realistic device modeling from atomic detail to the nanoscale
}


\author{J. E. Fonseca         \and
        T. Kubis \and
        M. Povolotskyi \and
        B. Novakovic \and
        A. Ajoy \and
        G. Hegde \and
        H. Ilatikhameneh \and
        Z. Jiang \and
        P. Sengupta \and
        Y. Tan \and
        G. Klimeck 
}

\authorrunning{J. E. Fonseca et al.} 

\institute{J. E. Fonseca \at
              Network for Computational Nanotechnology
              Purdue University
              West Lafayette, Indiana, US
              Tel.: 1765-496-6495\\
              \email{jfonseca@purdue.edu}           
           \and
           G. Klimeck \at
               \email{gekco@purdue.edu}
}

\date{Received: date / Accepted: date}

\maketitle

\begin{abstract}
As semiconductor devices scale to new dimensions, the materials and designs become more dependent on atomic details. NEMO5 is a nanoelectronics modeling package designed for comprehending the critical multi-scale, multi-physics phenomena through efficient computational approaches and quantitatively modeling new generations of nanoelectronic devices as well as predicting novel device architectures and phenomena. This article seeks to provide updates on the current status of the tool and new functionality, including advances in quantum transport simulations and with materials such as metals, topological insulators, and piezoelectrics.

\keywords{nanoelectronics \and Greens function formalism
(NEGF) \and NEMO \and tight-binding \and quantum dot \and strain \and transport\ and phonons \and Poisson \and parallel computing}

\end{abstract}

\section{Introduction}
\label{intro}

Relentless downscaling of transistor size has continued according to Moore’s law for the past 40 years. Transistor size will continue to decrease in the next ten years, but foundational issues with currently unknown technology approaches must be pursued \cite{2012}. This downscaling has reached the range where the number of atoms in critical dimensions is countable, geometries are formed in three dimensions and new materials are being introduced. Under these conditions we argue that the overall geometry constitutes a new material that cannot be found as such in nature \cite{Fuechsle2012}. Quantum effects such as tunneling, state quantization, and atomistic disorder dominate the characteristics of these nano-scale devices. 

The interactions of electrons, photons, and phonons are now governed by these new material properties and long-range interactions such as strain and gate fields. The end-game of the transistor size down-scaling as we know it is now fundamentally in sight. The end-game transistor is expected to be about 5nm long and 1nm in its critical active region corresponding to about 5 atoms in width. The physical atomistic down-scaling limit will be reached in about 8-10 years. The overall agenda is to bridge \emph{ab initio} materials science into TCAD simulations of realistically large scaled devices and get macroscopic quantities like current, voltages, absorption, etc., by mapping \emph{ab initio} into basis sets of lower order and include them in a formalism that allows for transport. The NEMO5 nanoelectronics modeling software is aimed at comprehending the critical multi-scale, multi-physics phenomena and delivering results to engineers, scientists, and students through efficient computational approaches and quantitatively modelling new generations of nanoelectronic devices in industry, as well as predicting novel device architectures and phenomena. 

\begin{sloppypar}
The basic functionality and history of the NEMO tool suite has been discussed previously \cite{Steiger2011,Sellier2012}. NEMO5's general software framework can easily include any kind of atomistic model and even semi-classical models if necessary. The scalable software implements Schr\"{o}\-dinger's equation and non-equilibrium Green's function method (NEGF) in tight-binding formalism, for electronic structure and transport calculations, respectively. It also is able to take into account important effects such as atomistic strain, using valence force field (VFF) strain models. It then allows the calculation of electronic band structures, charge density, current and potential, eigen-energies and wave-functions, phonon spectra, etc., for a large variety of semiconductor materials and devices. 

This manuscript seeks to provide interested readers with an overview of the recent developments surrounding NEMO5. This paper discusses approaches a) to quantum transport solutions, b) newly-implemented approaches to achieve faster convergence in the self-consistent Poisson-transport solution, c) strain, d) phonons, e) semi-automated material parameterization, f) metals, g) piezoelectric materials such as SmSe, h) topological insulators and i) band structure unfolding.
\end{sloppypar}
\section{Transport}
\label{transport}

\begin{sloppypar}
At the heart of NEMO5's quantum transport approach is the non-equilibrium Green's function method (NEGF) which is a computational approach to handle quantum transport in nanoelectronic devices \cite{Datta2005}. NEGF is numerically expensive when applied on atomistic tight-binding representations. NEGF requires storage, inversion and multiplication of matrices of the order of the number of electronic degrees of freedom. A well known method to ease the numerical burden is the recursive Green’s function method (RGF) that allows for limiting the calculation and storage of the retarded Green’s function to specific matrix blocks (such as only block diagonals and a single block column). Until recently, the RGF algorithm was limited to quasi 1D transport regimes, i.e. devices with 2 leads only. Generalizing work of Cauley et al., however, shows that RGF can be applied on virtually any transport problem, if the device Hamiltonian matrix is partitioned in a proper way \cite{Cauley2011}. NEMO5 allows partitioning the device ideally for 1D and quasi 1D transport problems according to the transport coordinate, but it also allows for the partition of complex, multi terminal devices and the application of RGF on them.

\begin{figure}[ptb]
  \includegraphics[width=0.48\textwidth]{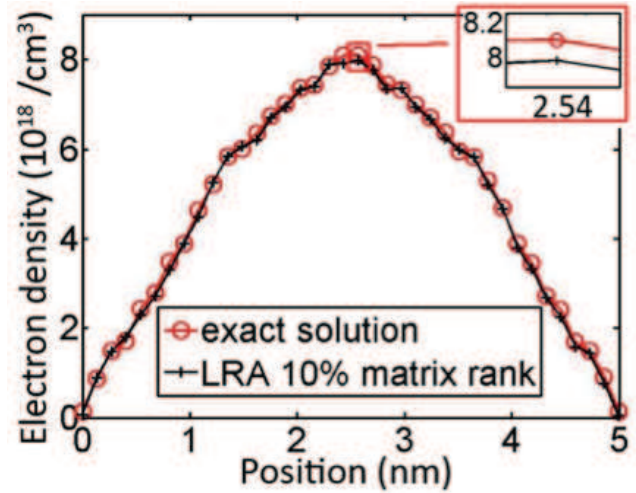}

\caption{ Comparison of the electron density of the exact NEGF calculation (circle) and of NEGF calculations with 10\% of the original matrix rank.}
\label{figLRA}       
\end{figure}

Despite the RGF method, the computational burden in memory and CPU time is still limiting the maximum device size solvable with NEGF. To overcome this obstacle, NEMO5 offers incomplete spectral transformations of NEGF equations into a Hilbert space of smaller rank than the original tight-binding representation \cite{Zeng2013}. Special cases of this low rank approximation are known as CBR method (all ballistic NEGF) \cite{Mamaluy2003} and the mode space approach \cite{Wang2004}. This method allows approximating NEGF transport problems in electronic tight-binding representations within a fraction of the numerical load of exact NEGF solutions. The loss of the NEGF accuracy and predictive power is thereby negligible as shown in Fig. ~\ref{figLRA} This figure compares the conduction band electron density of a homogeneous 5x5nm Si nanowire in equilibrium calculated in an exact and a LRA-approximate NEGF calculation where the rank has been reduced down to 10\% of the original problem size. Negligible discrepancies are magnified in the figure's inset.

Purely ballistic charge transport can be well described within the quantum transmitting boundary method (QTBM) \cite{Ting1992}. Since this method solves the quantum transport in the space of propagating lead modes, the numerical load is typically much smaller than in ballistic NEGF or RGF calculations which in general consider all modes. NEMO5 is able to solve the QTBM equations spatially distributed over large numbers of CPUs. For a given energy and transverse momentum, the boundary equations of the source and the drain are solved each on individual CPUs, whereas those sections of the device that are not in direct contact with the leads are solved on the remaining CPUs.
\end{sloppypar}

\section{Self-consistent calculation}
The many-body problem is treated in the Hartree approximation by self-consistently solving the Poisson and transport equations (e.g. QTBM, as explained above). The self-consistent solution is a nonlinear problem and any efficient solution of this problem must take into account at least three components: the energy grid, the initial guess, and the self-consistent algorithm. The energy grid should resolve the features in the energy dependent device charge density, determined by the lead density of states and device transmission properties. A good energy grid should be inhomogeneous, so that it is able to resolve sharp features, yet have as few energy points as possible to facilitate efficient computations. Since the self-consistent process is necessarily iterative in nature, the initial guess is the first step in the solution. A good initial guess, close to the final solution, can prevent convergence problems. The self-consistent algorithm provides the next potential guess in each iteration. Ideally, the algorithm should prevent divergence and arrive at the solution with as few as possible iterations. Via PETSc, NEMO5 employs several kinds of Newton-Raphson algorithms \cite{Balay2013a}, that rely on an efficient and approximate Jacobian implementation \cite{Trellakis1997,Lake1997} and have protection against divergence by being able to control the potential update, or step size, between two iterations. The Newton-Broyden method and trust region methods \cite{Cowell1984} are also used. While robust, these methods do not always guarantee efficient solutions. We achieve the most efficient solutions by constructing an accurate and time-efficient initial guess, based on the semi-classical charge and locally constant Fermi level with the effective mass corrected for confinement effects, followed by the Newton-Raphson method with full step size.

\begin{figure}[ptb]
\includegraphics[width=0.48\textwidth]{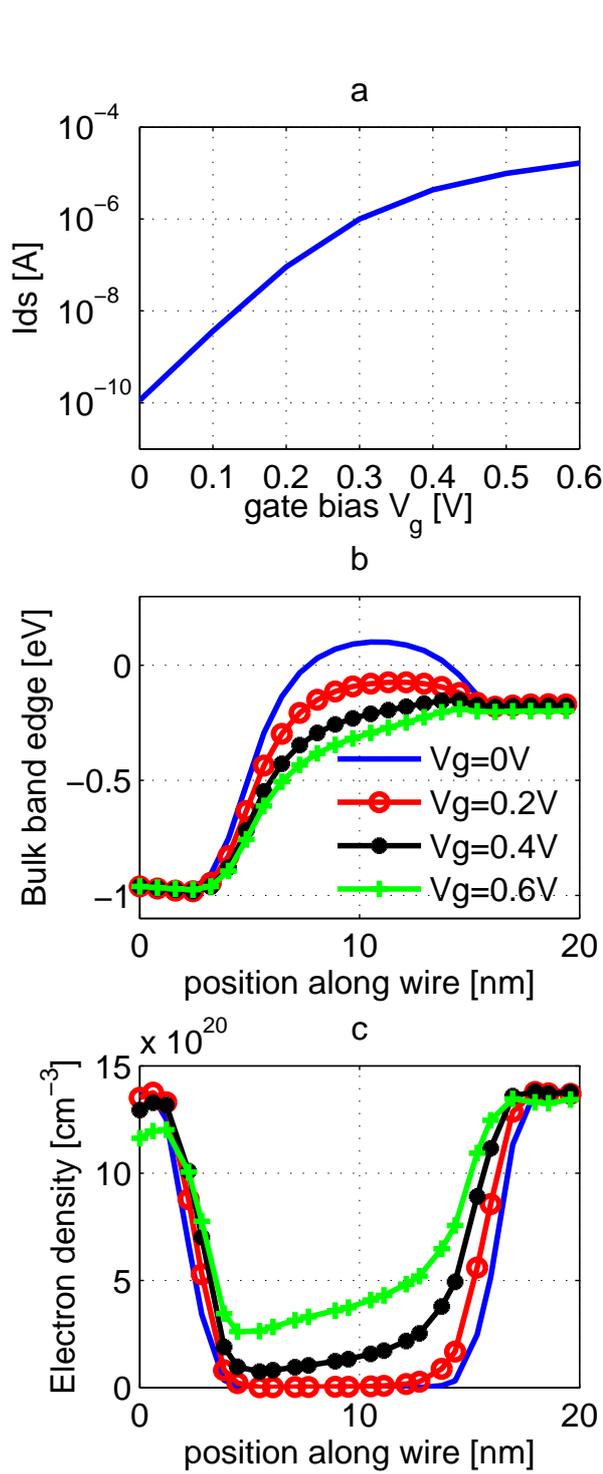}
\caption{NEMO5 self-consistent simulation results for n-type all-around gate Si nanowire. The gate length is 10 nm, while the doping in the channel below the gate is 10$^{15}$ cm$^{-3}$. The source and drain regions are taken to be 5 nm with 10$^{20}$ cm$^{-3}$ doping. Panel a) shows the current-voltage characteristic, b) is the bulk band edge interpolated along the center of the nanowire, and c) is the same for electron density. The charge density is nonuniform in the cross section due to lateral quantum confinement and significantly larger than the converged average charge which equals the doping. The simulation had 7 voltage points and took in total 27 Poisson/transport equation iterations, thanks to an efficient convergence scheme.}
\label{figSinanowire}       
\end{figure}

Results of one self-consistent simulation using the tight-binding formalism in NEMO5 are shown in Fig. ~\ref{figSinanowire}. The simulated device is n-type Si nanowire with 3x3 nm cross section (approximately 3 nm). The wire has 1 nm thick gate all-around and three doping regions: the channel under the gate is doped to 10$^{15}$ cm$^{-3}$, while the source and drain regions to 10$^{20}$ cm$^{-3}$. The length of the simulated device is 20 nm, of which 10 nm is the channel and 5 nm the source and drain regions each. The source and drain region length is chosen so that the potential becomes flat near the lead-device interface. Results are shown in Fig. ~\ref{figSinanowire}. The simulation is performed up to 0.6 V gate bias, to avoid unphysical effects at higher bias produced by ballistic transport in the absence of the transport barrier and subsequent Poisson/transport equation convergence issues. The convergence scheme consists of the Newton-Raphson method with full step size and the following initial guesses: for the first bias point the semi-classical initial guess is used; for the second bias point the previous solution is used as the initial guess; and for the third bias point upward the prediction/extrapolation based on the previous two solutions. This convergence scheme takes a total of 27 iterations for the 7 bias points simulated. The majority of inner bias points took only 3 iterations, while the first and the last bias point resulted in a slightly higher number of iterations. Even though the semi-classical initial guess with the effective mass corrected for confinement effects is very close to the final solution, the fact that the spatial effects of the confinement (i.e. quantum wave function) are not taken into account results in slightly more iterations. On the other hand, the quality of the semi-classical guess protects the simulation from divergence, as the full step size is used. The last bias point takes slightly more iterations, due to the fact that it is more difficult to achieve convergence for diminishing transport barrier at high gate bias.

\section{Strain}
\label{strain}

\begin{figure}[ptb]
\includegraphics[width=0.48\textwidth]{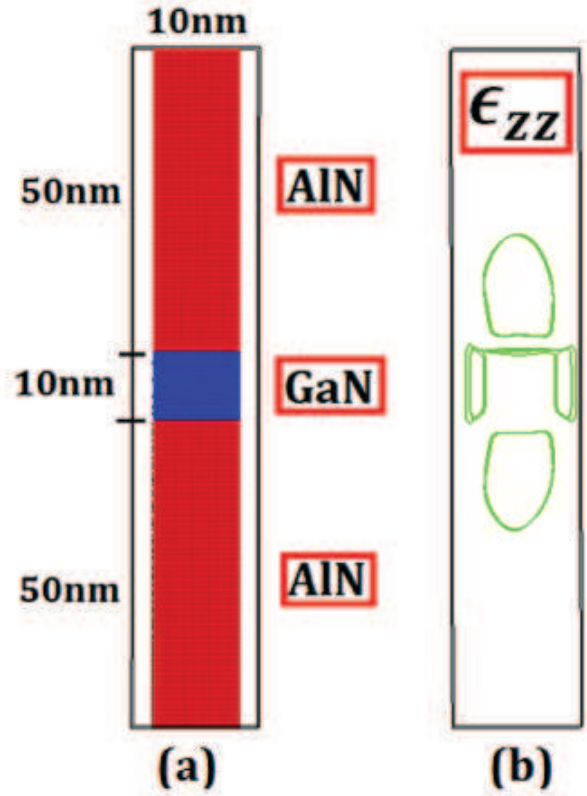}
\caption{Strain simulation in a Nitride Hetero-structure Nanowire using NEMO5. (a) Physical structure and dimensions, (b) plot of strain component ezz which shows long range diffusion of strain.}
\label{figstrain1}       
\end{figure}

\begin{sloppypar}
In the last decade, strain was a major performance booster in ultra-scaled transistors \cite{Antoniadis2006} and it is of fundamental importance to consider the effect of strain on the band-structure and transport properties of novel devices. Heterostructures composed of lattice mismatched materials exhibit strain intrinsically. As shown in Fig.~\ref{figstrain1} NEMO5 is able to compute strain and relax the atomistic heterostructures using the Enhanced Valence Force Field (EVFF) \cite{Steiger2011a,Paul2010,Sui1993}. The energy functional contains not only Keating terms such as bond-stretching and bond-bending interactions, but also cross-stretching, stretch-bending, and second-nearest-neighbor angle-angle interactions. For polar materials, the long-range Coulomb interaction can be added in the case of 0-D (bulk) and 3-D (confined) simulations. NEMO5 contains two strategies for elastic energy minimization. One uses Jacobian and Hessian matrices and can be used only for small structures. The second method is approximate and is based only on the Jacobian. 
\end{sloppypar}

\section{Phonons}
\label{phonons}

\begin{figure}
\includegraphics[width=0.48\textwidth]{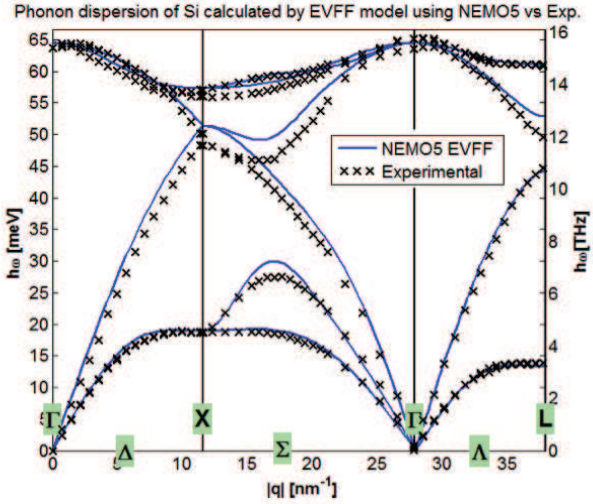}
\caption{Phonon dispersion of Si calculated by EVFF model using NEMO5 vs. experiment.}
\label{figphonon}       
\end{figure}

\begin{sloppypar}
Nanowires show excellent thermo-electric properties which make them favorable for thermo-electric devices. For example silicon nanowires exhibit 100 times better ZT compared to bulk silicon and can achieve maximum ZT around 1 \cite{Boukai2008}, creating a strong motivation for accurate phonon modeling in nanoscale devices. It is well known that the Keating model overstimates phonon energies of both optical and acoustic branches \cite{Sui1993}. NEMO5 is able to calculate phonon dispersion using the EVFF model which provides a reasonable match with experimental phonon dispersion (Fig.~\ref{figphonon}).
The dynamical matrix has been calculated by the following:
\end{sloppypar}
\begin{equation}
D_{\lambda,\mu}^{i,j} = \frac{1}{\sqrt{M_{i}M_{j}}} \frac{\partial^{2}U}{\partial r_{i}^{\lambda }\partial r_{j}^{\mu}}e^{-i\vec{q}\cdot \vec{r_{ij}}}   
\end{equation}

in which i and j are atom indexes, $\lambda$ and $\mu$ can be one of x, y or z directions, \em${q}$\em is the phonon wave vector, M$_{i}$ and M$_{j}$ are atom masses for atom i and j respectively and U is the total elastic energy of the system. 

\section{Metal insulator transition - SmSe}

With shrinking physical dimensions, the total transistor number in a single chip has been increasing exponentially for each generation. However, the scaling of the supply voltage in Silicon based MOSFET is limited by the 60mV/dec subthreshold swing (SS). The desire to reduce heat dissipation drives research for devices with different switching mechanisms. \cite{Theis2010}

\begin{figure}

\includegraphics[width=0.48\textwidth]{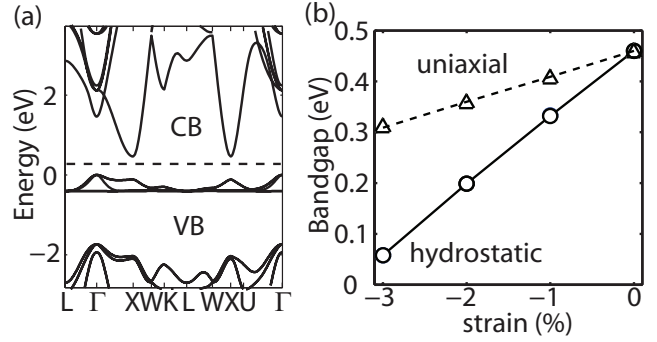}
\caption{Band structure of SmSe calculated with ETB. (a) bulk band structure of SmSe. (b) modification of bandgap under hydrostatic and uniaxial strain.}
\label{figSmSe}       
\end{figure}

The Piezoelectronic Transistor (PET) \cite{Newns2012,Newns2012a} is a promising approach to achieve a high ON/OFF ratio with very small voltage swing. In PET, the gate voltage is transduced to acoustic waves through a buffer layer made with piezoelectric (PE) materials. The channel layer of piezoresistive (PR) materials, e.g. Samarium monochalcogenides, is capable of modifying the conductance by several orders of magnitude subjected to moderate strain \cite{Mott1968} which is generated by deformation of PE. 
When the dimensions of PET are reduced to the nanometer scale, the device performance will be dominated by quantum effects. Quantum confinement will change band structure and minimum leakage is determined by tunneling. To simulate devices of realistic dimensions, computationally efficient models like Empirical Tight-Binding (ETB) are necessary. 
To obtain accurate parameterization, the SmSe band structure was first calculated in density functional theory (DFT) within the generalized gradient approximation with Hubbard-type U (GGA+U). A tight-binding band model including {\em spdfs*} orbitals is implemented based on analysis of the DFT angular momentum decomposition at the band minima \cite{Tan2013,Jiang2013}. The inclusion of enhanced spin-orbit coupling for {\em f}-orbit is critical to account for the large {\em 4f$_{5/2}$-4f$_{7/2}$} splitting due to a strong electron-electron interaction of localized {\em f} electrons. This model captures the band structure features and the variations of the bandgap in response to the strain predicted by DFT calculations (Fig.~\ref{figSmSe}). The obtained TB parameters are then used in quantum transport simulations with (NEGF). 

\section{Material parameterization}
\label{yaohua}

The ETB method is widely used in atomistic device simulations. The reliability of such simulations depends very strongly on the choice of basis sets and the ETB parameters. The traditional way of obtaining the ETB parameters is by fitting to experiment data, or critical theoretical bandedges and symmetries rather than a foundational mapping. A further shortcoming of traditional ETB is the lack of an explicit basis.

\begin{figure}
\includegraphics[width=0.48\textwidth]{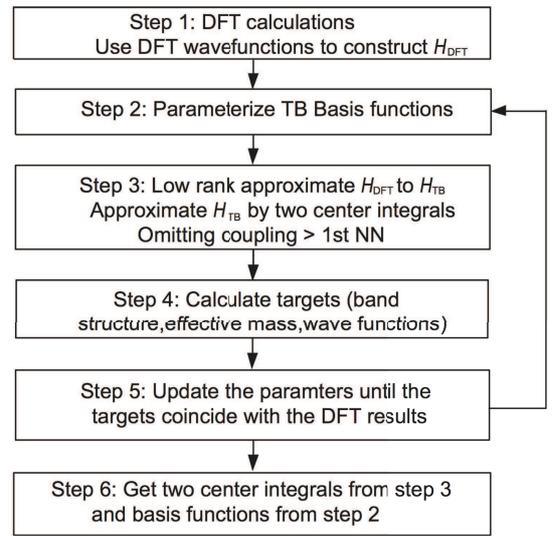}
\caption{The process of tight-binding (TB) parameters construction from DFT calculations.} 
\label{figflowchart}
\end{figure}

The mapping method is described here is shown in Fig.~\ref{figflowchart}. The first step is to perform ab-initio calculations of the band structure of a material. In general, any method that is capable to calculate electronic band structures and wave functions is suitable here. In the second step, the ETB basis functions for each type of atom are defined as

\begin{equation}
\Psi _{n,l,m}\left( \mathbf{r}\right) \equiv \Psi _{n,l,m}\left(
r,\theta ,\phi \right) =R_{n,l}\left( r\right) Y_{l,m}\left( \theta
,\phi \right) , \label{eq:definition_atomic_orbitals}
\end{equation}%
where the functions $Y_{l,m}$ are the complex spherical harmonics
with angular quantum numbers $l$ and $m$; and the functions $R_{n,l}$ are exponentially damped plane waves%
\begin{equation}
R_{n,l}\left( r\right) =\sum_{i=1}^{N}\left[ a_{i}\sin \left( {\lambda _{i}r}%
\right) {+b_{i}\cos }\left( {{\lambda _{i}r}}\right) \right]
r^{n-1}\exp \left( -\alpha _{i}r\right) .
\label{eq:definition_basis_Radialpart}
\end{equation}%
The parameters $a_{i},b_{i},\alpha _{i},\lambda _{i}$ are the fitting parameters. With a given set of ETB basis functions $\Psi _{\text{TB} }^{\mathbf{k}}$, the DFT Hamiltonian is transformed to the tight-binding representation. Any non-zero off-diagonal element of the overlap matrix is neglected. The ETB Hamilton matrix elements are approximated by two center integrals according to the Slater-Koster table~\cite{Slater_Tightbinding,Podolskiy_TBElements}. ETB Hamilton matrix elements beyond either 1st or 2nd nearest neighbor coupling are neglected. In Step 4, the band edges, effective masses and eigen functions of the Hamiltonian at high symmetry points are calculated and compared to the corresponding DFT results. The overlaps of the ETB basis functions are also determined. In the fifth step, all fitting parameters are adjusted to improve the agreement of the ETB\ results with the DFT results and also to reduce the overlap matrix of the ETB basis functions to the unity matrix. Steps 2 - 5 are repeated until the convergence criterion is met. Step 6 requires to extract the converged ETB basis functions and the ETB two center integrals.
\begin{figure}[t]
\includegraphics[width=0.48\textwidth]{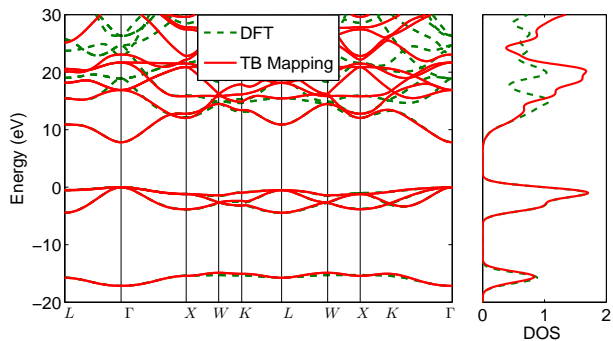}
\caption{Band structure and density of states of MgO by DFT and TB.}
\label{figfig_Ek_TE_MgO}
\end{figure}

The DFT mapping method has been validated in NEMO5 using Si and GaAs\cite{Tan2013}. It is also applied successfully to Antimonides and new materials such as MgO\cite{Tan2013}, SmSe\cite{Jiang2013}. MgO cyrstalizes in rock salt structure. Each oxygen atom has six magnesium atoms as 1st nearest neighbors and twelve oxygen atoms as 2nd nearest neighbors. MgO is parameterized for a 2nd nearest neighbor $sp3d5s^*$ ETB model. Within this model, the interaction between two oxygen atoms is required to produce the correct valence bands. It can be seen from Fig.\ref{figfig_Ek_TE_MgO} that the ETB band structure matches the DFT result well within the energies $-5$ to $15~\mathrm{eV}$.

\section{Tight-binding parameterization of metals}
\label{metals}

\begin{sloppypar}
Metals play a significant role in microprocessor operation. Besides connecting individual transistors, metal vias and interconnects deliver bias current, clock signals and metal-stacks are used for gate metallization. In the existing quantum mechanical atomistic device-modeling paradigm, metal contacts are used to set the Fermi levels in the source and drain of the transistor. Once this is done, the metals are essentially abstracted out of the usual Schr\"{o}dinger-Poisson or NEGF-Poisson solution. Owing to decreasing device dimensions, the resistance drop across the metal-semiconductor contact is becoming an increasingly important issue. Additionally, from an overall power dissipation perspective, the increase in metal resistivity with decreasing via dimensionality is an extremely important unsolved problem\cite{2012}. Atomistic modeling of metal grain boundary interfaces, metal interconnect-liner interfaces and metal-semiconductor interfaces can provide significant guidance in the design of low-resistivity metal interconnects, liner materials and metal-semiconductor interfaces with low Schottky-barrier heights. With these objectives in mind, we have created accurate and computationally efficient Semi-Empirical Tight-Binding (SETB) models of Metals and Metal-Semiconductor interfaces suitable for studying electron transport in the aforementioned, technologically important systems.

\begin{figure}
\includegraphics[width=0.48\textwidth]{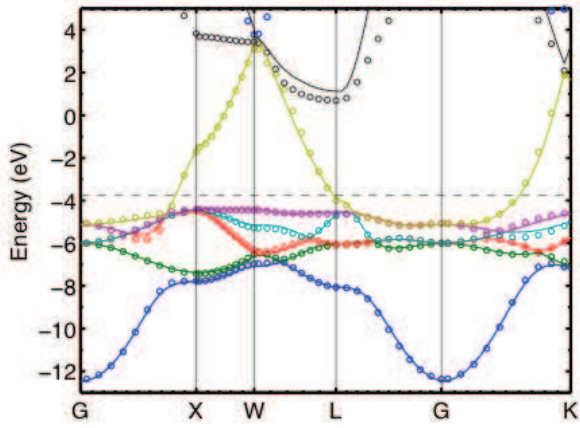}
\caption{Bulk band structure of Cu in FCC phase calculated using the SETB (dots) formalism and DFT (solid lines). Notice that our SETB model accurately captures DFT band structure features in all energies of interest in electronic transport. 
}
\label{figmetalsa}       
\end{figure}

\begin{figure}
\includegraphics[width=0.48\textwidth]{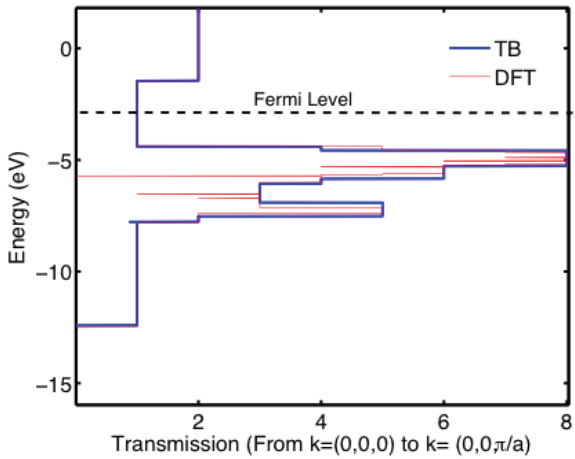}
\caption{Transmission for a 1 nm cell along the [001] direction in Cu computed using SETB and DFT.
}
\label{figmetalsb}       
\end{figure}

NEMO5 contains tight-binding models that have been formulated specifically to study the phenomenon of resistivity increase in metals with decreasing interconnect dimensions and electron transport across metal-semiconductor interfaces. As an example of the capabilities NEMO5 has in this regard, Fig.~\ref{figmetalsa} shows the bulk band structure of Cu obtained using an efficient 1st Nearest-Neighbor SETB representation of its FCC phase. This band structure is obtained by fitting to LCAO Density Functional Theory (DFT) band-structure for Cu using the exchange-correlation functional of Perdew and Zunger within the Generalized Gradient Approximation (GGA)\cite{Perdew1996}. It can be seen that the tight-binding model reproduces the DFT band structure accurately. In Fig.~\ref{figmetalsb}, the transmission in bulk Cu along the [001] direction is computed using SETB and DFT and the results are compared to each other. It is evident that our SETB model reproduces the DFT results extremely accurately in the energy range of interest – a few kT’s below and above the Fermi level. 
\end{sloppypar}

\section{Topological insulators - Bi$_{2}$Te$_{3}$}
\label{BiTe}
\begin{figure}
\centering
\includegraphics[width=0.48\textwidth]{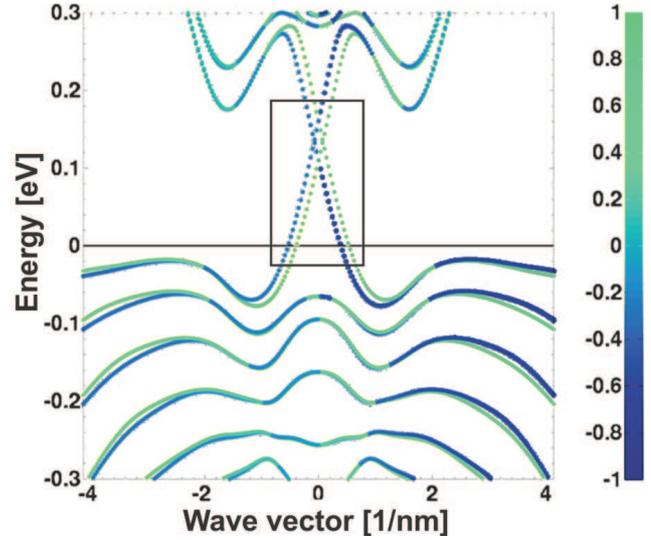}
\caption{The bandstructure of a $[100]$ grown Bi$_{2}$Te$_{3}$ quantum well. The surface states are shown within the boxed region. The conduction and valence bands are connected by a linear dispersion also known as a Dirac cone, depicted within the box. The color bar denotes the strength of spin-polarization.}
\label{figBiTespinprojection}
\end{figure}

\begin{sloppypar}
Topological Insulators (TIs) are a new state of matter with a bulk insulating gap and metal-like states on the surface or edge. The surface or edge states which are described by a linear Dirac Hamiltonian are robust and protected by time reversal symmetry \cite{hasanrmp,fuprb07}. Topological insulators have acquired prominence because they offer a rich collection of fundamentally new phenomena along with a wide array of applications including optoelectronic THz detectors, spin-polarized contacts, ultra-fast switches, etc. \cite{Sengupta2013}. Several TI materials are known to exist at room temperature. Bi$_{2}$Te$_{3}$ and Bi$_{2}$Se$_{3}$ which possess bound surface states (Fig.~\ref{figBiTespinprojection}) are well-known examples.

The unique properties of TIs are attributed to the linear dispersion of surface states that connect the conduction and valence band together. Further, these states have their spin locked perpendicular to momentum in-plane. NEMO5 offers the capability to compute the atomistic band structure of bulk and confined Bi$_{2}$Te$_{3}$ devices. The undoped Bi$_{2}$Te$_{3}$ is a narrow band-gap quintuple-layered semi-conductor with a rhombohedral crystal structure. The quintuple layer crystal structure is used in a twenty band tight-binding model. All parameters for these calculations were obtained from a orthogonal tight-binding model with sp$^{3}$d$^{5}$s* orbitals, nearest-neighbor interactions, and spin-orbit coupling \cite{lee2006tight}. Additionally, the dispersion is spin-resolved and conforms exactly to experimentally observed spin-polarization (see Fig.~\ref{figBiTespinprojection}).
\end{sloppypar}

\begin{figure}
\centering
\includegraphics[width=0.48\textwidth]{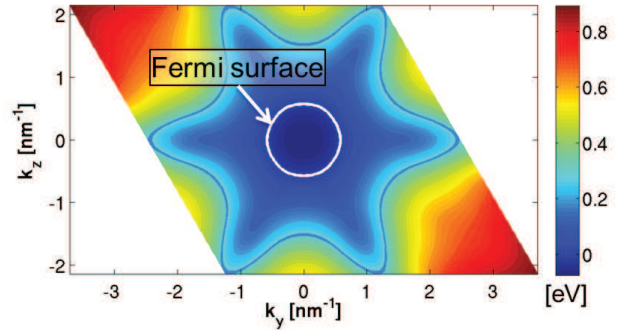}
\caption{The Fermi-surface of the surface states of Bi$_{2}$Te$_{3}$ with the distinctive snow-flake structure.}
\label{figFermi}
\end{figure}

The corresponding Fermi-surface of the surface states exhibit a peculiar snow-flake structure. NEMO5 predicts this (Fig.~\ref{figFermi}), in agreement with experiments ~\cite{souma2011direct}.

When conduction band and valence bands are connected (as in the case of TIs) conduction and valence bands cannot be unambiguously separated. Since an accurate prediction of device characteristics and material properties needs charge self-consistent calculations, it is imperative to devise electronic structure calculation models for materials where an explicit differentiation between electrons and holes is not possible. NEMO5 introduces the concept of a novel charge self-consistent full-band atomistic tight-binding method that avoids usage of holes.  Hereby, the model of Andlauer and Vogl has been extended to atomistic tight-binding \cite{Andlauer2009}.

\section{Bandstructure unfolding}
\label{unfolding}

\begin{sloppypar}
Semiconductor alloys do not possess translational symmetry, owing to a random distribution of atoms. For example, the cationic sites in Al$_{x}$Ga$_{1-x}$As can either accommodate an Al or a Ga atom. Thus semiconductor alloys cannot, in principle, have an associated bandstructure. Nevertheless, it is common to measure and use quantities associated with bandstructure (for example, energy bandgap and effective mass) to design and analyze devices in these materials. A compromise between the above two positions is to allow for an approximate bandstructure of alloys, where each energy band is broadened as a result of randomness. The supercell method \cite{Boykin_JPhysCondMatter_2007,Popescu_PRB_2012} provides a computational framework to perform such a calculation. The essential idea is to construct a very large supercell which is randomly populated with atoms. A supercell of say Si$_{0.4}$Ge$_{0.6}$ would have roughly $40 \%$ of atoms being Si, while the rest being Ge. Periodic boundary conditions are imposed on this large supercell, and its energy spectrum determined (typically at a single $\vec{K}$ point). The supercell is viewed as being made up of fictitious primitive cells called \emph{small-cells}. The supercell energy spectrum is finally unfolded onto the small-cell Brillouin zone and approximate small-cell energy bands determined.
\end{sloppypar}

\begin{figure}
 \centering	
 \includegraphics[scale=0.8]{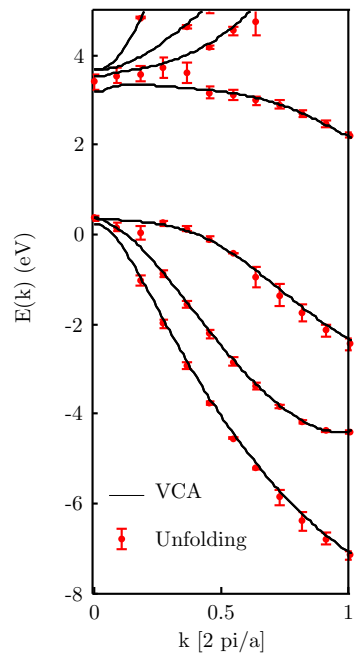}
 \caption{Energy bands of Si$_{0.5}$Ge$_{0.5}$ alloy obtained using 
the supercell method compared with those obtained with the VCA method.}
\label{fig_SiGe110}
\end{figure}

In order to obtain adequate points along a particular direction $\vec{n}$ in the small-cell Brillouin zone (say $[100]$, $[110]$, $[111]$ etc.), it is convenient to work with specially chosen supercells. Reference \cite{Boykin_Physica_2009} describes special rectangular, non-primitive unitcells that are used as building blocks to construct such supercells in NEMO3D \cite{Klimeck_TED_2007A,Klimeck_TED_2007B}. This approach has two drawbacks -- (i) the non-primitive unit cell is itself made up of a number of small cells, requiring an additional unfolding step that is dependent on $\vec{n}$; (ii) it cannot be used for materials (like GaN) which do not have rectangular unit cells. NEMO5 implements a more general approach, based on \cite{Aravind_AJP_2006,Ajoy_IWPSD_2012} where the supercell is built by cascading specially chosen primitive cells (which could be non-rectangular).
\begin{sloppypar}
Figure \ref{fig_SiGe110} shows the approximate energy bands of Si$_{0.5}$Ge$_{0.5}$ along the $[110]$ direction obtained by unfolding from a supercell containing 248 atoms. The atomic positions have been relaxed using a Keating model. Also shown are results of a virtual-crystal-approximation (VCA) \cite{Paul_EDL_2010}, which computes energy bands using a primitive cell consisting of virtual atoms, whose properties are obtained by interpolating those of Si and Ge. It is interesting to note that the VCA approach provides a good estimate of the energy bands of bulk SiGe; nevertheless, the VCA approach has been known to be erroneous for SiGe wires \cite{Klimeck_TED_2007B}. 
\end{sloppypar}

\section{Conclusion}
\label{Conclusion}
An overview of the the NEMO5 nanoelectronics modeling tool has been given with updates regarding recent advances in physical models and associated code. With focus on efficient, scalable quantum transport algorithms, combined with flexibility to handle a wide variety of device structure and materials, NEMO5 seeks to be a cohesive package to provide accurate modeling of nanoscale devices.



%
%

\begin{acknowledgements}
This work was partially supported by NSF PetaApps grant number OCI-0749140, NSF grant EEC-0228390 that funds the Network for Computational Nanotechnology, and SRC NEMO5 development: Semiconductor Research Corporation (SRC) (Task 2141), and Intel Corp.

With kind permission from Springer Science+Business Media: Journal of Computational Electronics, Empirical tight binding parameters for GaAs and MgO with explicit basis through DFT mapping, volume 12, issue 1, 2013, pages 56-60, Yaohua Tan, Michael Povolotskyi, Tillmann Kubis, Yu He, Zhengping Jiang, Gerhard Klimeck, and Timothy B. Boykin, figures 6 and 7, \copyright Springer Science+Business Media New York 2013 DOI 10.1007/s10825-013-0436-0.
\begin{sloppypar}
The final publication the in the Journal of Computational Electronics is available at http://link.springer.com/article/10.1007\%2Fs10825-013-0509-0
\end{sloppypar}
\end{acknowledgements}

\bibliographystyle{unsrt}
\bibliography{References_N5BiTe,exportlistkeys,refs_Unfolding,apssamp}   

%
%

\end{document}